\documentclass[twocolumn,preprintnumbers,showpacs,aps,prb]{revtex4}

\usepackage{graphicx,times,epsfig,color}
\usepackage{float}

\usepackage{comment}

\begin{document}

\def\salto{\vskip 1cm}
\def\lgr{\langle\langle}
\def\rgr{\rangle\rangle}

\title{Competition between Kondo effect and RKKY physics in graphene magnetism}
%
\author{A. Allerdt}
\affiliation{Department of Physics, Northeastern University, Boston, Massachusetts 02115, USA}
\author{A. E. Feiguin}
\affiliation{Department of Physics, Northeastern University, Boston, Massachusetts 02115, USA}
\author{S. Das Sarma}
\affiliation{Department of Physics, Condensed Matter Theory Center and Joint Quantum Institute, University of Maryland, College Park, Maryland 20742-4111, USA}

\begin{abstract}
The cooperative behavior of quantum impurities on 2D materials, such as graphene and bilayer graphene, is characterized by a non-trivial competition between screening (Kondo effect), and Ruderman-Kittel-Kasuya-Yosida (RKKY) magnetism. In addition, due to the small density of states at the Fermi level, impurities may not couple to the conduction electrons at all, behaving as free moments. 
Employing a recently developed {\em{exact}} numerical method to study multi-impurity lattice systems, we obtain non-perturbative results that dramatically depart from expectations based on the conventional RKKY theory. 
At half-filling and for weak coupling, impurities remain in the local moment regime when they are on opposite sublattices, up to a critical value of the interactions when they start coupling anti-ferromagnetically with correlations that decay very slowly with inter-impurity distance. 
At finite doping, away from half-filling, ferromagnetism is completely absent and the physics is dominated by a competition between anti-ferromagnetism and Kondo effect. In bilayer graphene, impurities on opposite layers behave as free moments, unless the interaction is of the order of the hopping or larger. 
\end{abstract}

\maketitle

\section{Introduction}
Over a decade after the first successful experimental realization of a truly two-dimensional material, graphene\cite{Novoselov.Electric_field_atomically_thin_carbon,graphene1b,geim2007}, interest in its properties continues to grow, 
with potential for future applications in data storage, spintronics, sensors, magnetic imaging, and quantum computing, to mention some examples\cite{Nalwa.MagneticNanostructures,avouris2007}. 

The physics of diluted magnetic impurities in graphene is rich and constitutes an entire subject of research in its own right \cite{castro2009adatoms,Lars, shytov2009long,Kotov2012}.
Isolated magnetic adatoms placed on mono-layer graphene sheets (MLG) have been studied experimentally as well as theoretically \cite{graphene_review1,graphene_review2},
and the properties of the Kondo ground state have been a subject of controversy 
 \cite{bulla1997anderson,Buxton.RG_anderson_gapless,uchoa2008localized,uchoa2011kondo,wehling2010,wehling2010theory,Lars,Withoff.Phase_transitions_gapless}. The Kondo effect due to magnetic
adatoms such as cobalt, presents different behaviors depending on the position of the impurities in graphene sublattices. In addition, orbital and magnetic moments of the impurities strongly depend on the used substrate \cite{Donati.Tailoring_magnetism_Co,Eelbo.Influence_of_decoupling, Eelbo.Adatoms_on_graphene}.
For adatoms directly on top of carbon sites, a Fermi liquid behavior consistent with an $SU(2)$ Kondo effect has
been predicted and found to be consistent with experimental results \cite{graphene_Kondo,cornaglia2009,sengupta2008,Jacob10}.
However, for adatoms at the center of a hexagon in the graphene lattice, the results are confusing and contradictory.
Based on symmetry arguments for MLG and bilayer graphene (BLG), four-channel and two-channel \cite{schneider2011,kharitonov2013kondo}, as well as an $SU(4)$ Kondo effect\cite{wehling2010} were 
predicted in the presence of spin-orbit coupling. 
Moreover, the Kondo state does not depend only on the position of the adatom, but also on the band filling or doping\cite{cornaglia2009}.
By gating graphene, one can move the Fermi energy $E_F$ (i.e. doping or band filling) away from the Dirac point to a region of the band with a linear density of states, in which case the Kondo effect becomes conventional. The important conceptual question addressed in this paper is how Kondo effect interplays with adatom-induced magnetism in MLG and BLG systems in the presence of multiple magnetic impurities in the graphene lattice.
We emphasize that the actual physics of graphene magnetism is complex and may depend on many details (e.g. precise locations of impurities, the strengths of their couplings to the bulk, their separations, the Fermi level, etc.), but the conceptually simple (and intuitively appealing) starting point of a perturbative RKKY-type inter-impurity magnetic interaction should always be treated with caution, and may not, in general, be applicable.

In this work we address the two impurity problem, which is usually treated by introducing an effective Ruderman-Kittel-Kasuya-Yosida (RKKY)\cite{Yoisida.Magnetic_properties,Kittel.Indirect_exchange,Kasuya.Theory_of_metallic} interaction between impurities, mediated by the conduction electrons in the system, derived from second-order perturbation theory:
\[
J_{RKKY}({\bf R})=J_K^2\chi({\bf R}),
\]
where $\chi({\bf R})$ is the Fourier transform of the non-interacting static susceptibility, or Lindhard function, and $J_K$ is the Kondo interaction between the impurity and substrate. The dependence of this function on the distance varies with dimensionality. A universal expression is often used in the literature, which is derived from assuming a uniform electron gas with a quadratic dispersion\cite{Aristov1997} $E(k) \sim k^2$. Its asymptotic behavior at long distances ($k_F R \gg 1$) and in $d$ dimensions is of the form:
\begin{equation}
\chi(R) \sim \frac{\sin{(2k_F R +\pi d/2)}}{R^d},
\label{lindhard}
\end{equation}
which can be ferro or anti-ferromagnetic (depending on $k_F$ and $R$), and oscillates with impurity separation $R$ and wave-vector $2k_F$ (twice the Fermi momentum). It is long-ranged with an amplitude that decays algebraically (in particular, as $1/R^2$ in two dimensions). 
This perturbative approach, however, fails to capture important many-body effects. It was previously shown\cite{Allerdt.Kondo} that geometry, band structure, and Kondo effect can drastically affect the physics here. For instance, in graphene, the RKKY interaction has contributions that decay both as $1/R^2$ and $1/R^3$, a reflection of the vanishing density of states (DOS) at the Fermi level in intrinsic ({\it i.e.} undoped) graphene \cite{ando2005,brey2007,Hwang2008}. 

For single magnetic impurities coupled to a metallic host, all physical properties can be characterized by a single energy scale, the so-called ``Kondo temperature'' $T_K\simeq e^{-1/J_K}$, which can interpreted as a binding energy for forming a Kondo singlet. When more impurities are present, several energy scales compete. As suggested in Ref.~\onlinecite{Doniach1977} (see also Refs.~\onlinecite{Schwabe2012,Schwabe2015}), one could define a characteristic temperature for coupling the impurities into the RKKY state, $T_{RKKY} \sim J_K^2$, and a competition between these two energy scales ($T_K$ and $T_{RKKY}$) will dictate which phase dominates. Moreover, in finite systems (Kondo box\cite{Thimm1999,schlottmann2001kondo,simon2002finite,simon2003kondo,hand2006spin,Hanl2014}) or in the presence of a gap, a third energy scale will enter into the problem: the level spacing, or gap $\Delta$. The impurities can potentially be found in three states: (i) They can couple via an effective indirect exchange interaction $J_{RKKY}$, (ii) they can each form their own independent Kondo singlets, or (iii) they can remain in a free moment state, completely decoupled from the substrate and from each other. These issues have not at all been discussed theoretically in the context of adatom-induced graphene magnetism. 

The paper is organized as follows: Section II describes the model and methods utilized, focusing on the Lanczos transformation to the equivalent one-dimensional model. We point out that this technique has been described in the literature in great detail \cite{Allerdt.Kondo,Yunoki.Block}, so we cover only the essential aspects. Section III describes the {\it exact} results obtained with this approach both for single, and bi-layer graphene, illustrating the departure from conventional perturbative predictions. We close our paper with a summary and conclusions.

\begin{figure}
\centering
\includegraphics[scale=0.28]{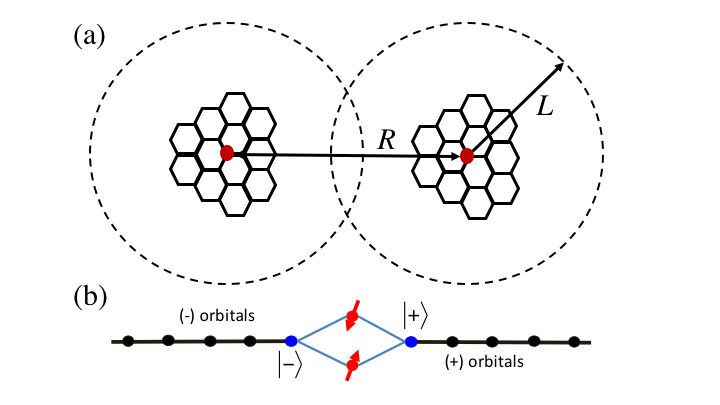}
\caption{
(a) Sketch of graphene flakes obtained through the Lanczos transformation for two impurities sitting at the origin, and at a distance $R$. The single particle orbitals extend to a distance $L$ away from the impurities. In (b) we show the geometry of the equivalent problem, with the two magnetic impurities coupled to non-interacting tight-binding chains of length $L$ via many-body terms proportional to $J_K$.
}
\label{orbitals}
\end{figure}

\section{Methods} 

In this work, we consider two $S=1/2$ magnetic impurities interacting locally with free fermions in the bulk.
In general, the total Hamiltonian of this problem can be written as:
\[
H = H_\mathrm{band} + H_\mathrm{imp} + V,
\]
where $H_\mathrm{band}$ is the lattice Hamiltonian, $H_\mathrm{imp}$ is the many body impurity Hamiltonian ({\it e.g.},
Coulomb interaction in the case of Anderson impurities), and $V$ contains the hybridization terms coupling the lattice and the impurities. 

We consider two models for the interaction. The first one involves a Kondo term between two spins at positions $r_1$ and $r_2$ and the substrate:
\begin{equation}
V = J_K \left( \vec{S}_1 \cdot \vec{s}_{\rm r_1} + \vec{S}_2 \cdot \vec{s}_{\rm r_2} \right).
\label{hamiltonian1}
\end{equation}
In a second setup we consider ad-atoms located at the center of hexagonal plaquettes, and
an Anderson-like impurity Hamiltonian of the form:
\begin{equation}
H_\mathrm{imp} = \epsilon_{d}\sum_{i=1,2}n_{di} + U\sum_{i=1,2}n_{di\uparrow}n_{di\downarrow}
\label{hamiltonian2}
\end{equation}
with
\begin{equation} 
V = t'\sum_{\sigma,i,\delta}d^\dagger_{i\sigma}c_{i_\delta\sigma}
\label{vterm}
\end{equation}
where $d$, $d^\dagger$, $n_d$ operators act on the two magnetic impurities $i=1,2$. In Eq.(\ref{vterm}), $\delta$ labels the six sites surrounding the impurities, and $\epsilon_d$ is the local site energy. The assumption that the impurity couples symmetrically to the six sites is valid if the atomic orbital has $s$, $d_{z^2}$, or $f_{z^3}$ symmetry \cite{Guessi.Catching}.
In both cases the models represent adatoms, and not substitutional impurities.

In order to make the problem numerically tractable, we employ the so-called block Lanczos method recently introduced in this context by two of the authors \cite{Allerdt.Kondo,Yunoki.Block}. This approach is inspired by Wilson's original formulation of the numerical renormalization group \cite{nrg_wilson}, but accounting for the lattice structure. It enables one to study quantum impurity problems in real space and in arbitrary dimensions with the density matrix renormalization group method (DMRG)\cite{White1992,White1993}. 
Our scheme bares resemblance to Haydock's recursion method\cite{Haydock1972,Haydock1975,Haydock1980,RecursionBook}, where the information about the lattice structure and the hybridization to the impurity is completely preserved.
By generalizing the ideas introduced in Ref.~\onlinecite{Feiguin.Lanczos} for a single impurity, one can reduce a complex lattice geometry to a single chain, or a multi-leg ladder in the case of multiple impurities.
This is done through a unitary transformation to a basis where the non-interacting band Hamiltonian has block diagonal form. 
As described in detail in Refs.~\onlinecite{Allerdt.Kondo,Yunoki.Block}, this is equivalent to a  block Lanczos iteration, where the recursion is started from seed states corresponding to electrons sitting at the positions of the impurities.
The resulting matrix can be re-interpreted as a single-particle Hamiltonian on a ladder geometry.

In addition, we use a folding symmetry transformation\cite{Allerdt.Kondo} that maps the ladder onto two decoupled chains corresponding to bonding and anti-bonding orbitals. This geometry is amenable to direct DMRG calculations, and allows one to simulate large one-dimensional systems. In a real space representation of the Lanczos orbitals, this corresponds to almost circular flakes of graphene of radius $L$, as shown schematically in Fig.~\ref{orbitals}. 

Notice that this canonical transformation is exact, and the only errors in our calculations can be attributed to finite-size effects (discussed in the next section), or numerical precision, which we keep under control.
Each impurity configuration generates a new mapping, and the equivalent one-dimensional problem is solved using the DMRG method, keeping the truncation error below $10^{-9}$, which translates into up to 3000 DMRG basis states. 
All calculations are done using a chain with length $L=4n$, which allows each impurity to form their own Kondo state or a collective RKKY state.\cite{Yanagisawa.Ground_state} Results shown were performed using a total chain length of $L=124$, so the system size is much larger than the maximum impurity separation. In all our simulations we use $U(1)$ symmetry to fix the total value of $S^z_\mathrm{Tot}=0$. 

\begin{figure}
\centering
\includegraphics[scale=0.4]{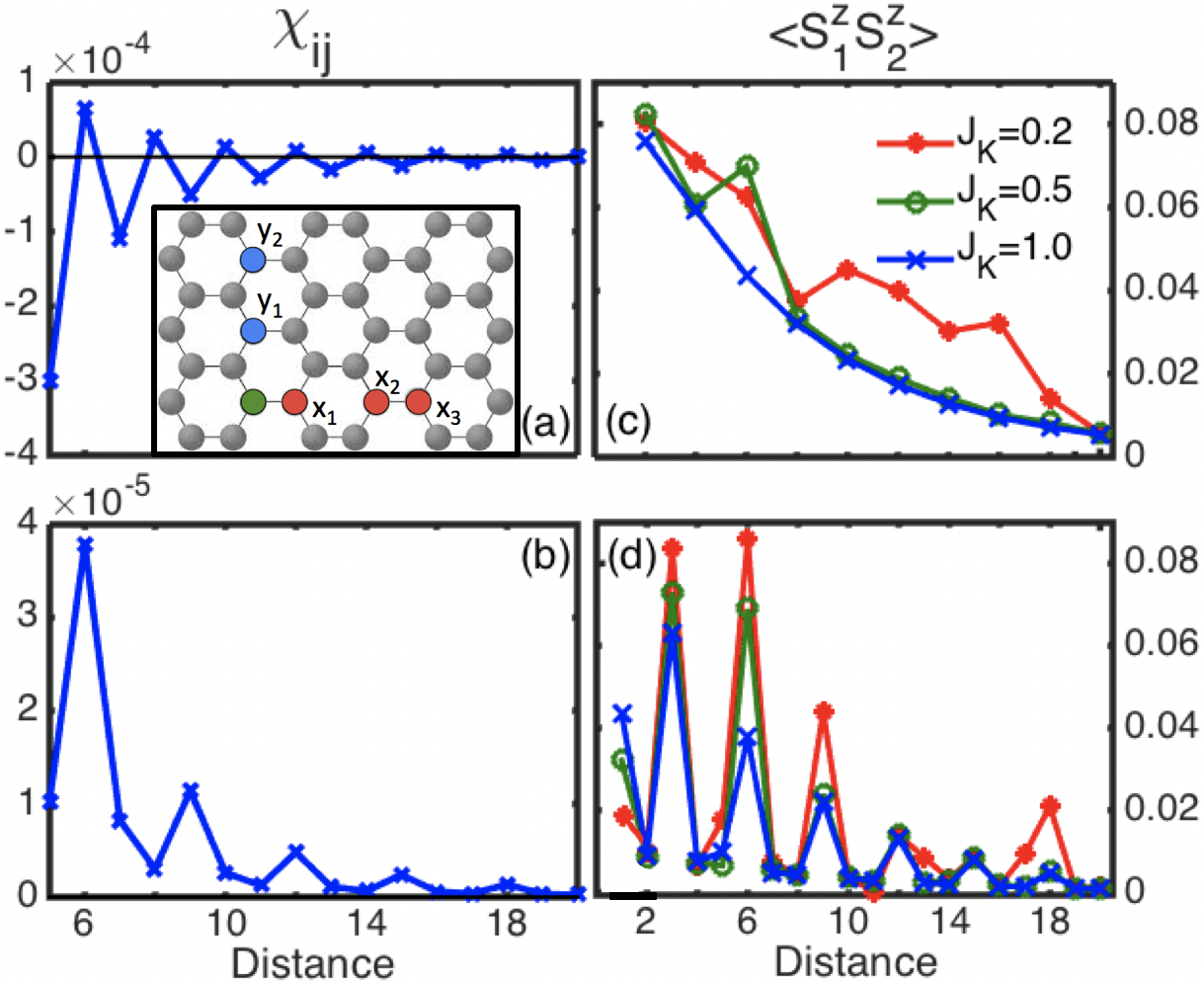}
\caption{Lindhard function for MLG at half-filling along the (a) $x$-direction and (b) $y$-direction, as depicted in the inset. Distances are not to scale and  label lattice sites, with the green circle representing the first impurity and the blue(red) circles representing the second impurity along the y(x) direction. Panels (c) and (d) show the non-perturbative results for the spin-spin correlations along the $x$ and $y$ directions, respectively. In (c), impurities on opposite sublattices are in the free moment regime or anti-ferromagnetically aligned and are shown in a separate figure.
}
\label{slg_hf}
\end{figure}

\section{Results}
\subsection{Graphene}
Graphene can be represented as a (bi-partite) honeycomb lattice of carbon atoms. It is well approximated using a tight-binding theory with only nearest neighbor hopping. We consider just one $p_z$ orbital per atom, giving us the simple two band symmetric model with a Dirac point at the Fermi level. In the following, our unit of energy is the hopping  $t\approx2.5eV$\cite{graphene_review2}.

As a reference, we first calculate the non-interacting Lindhard function, Eq. (\ref{lindhard}). The directions chosen, labeled $x$ and $y$, are shown in Fig.~\ref{slg_hf}(a) and (b). Due to the bipartite nature of the lattice, at half-filling the sign of the interaction oscillates and is ferromagnetic when the impurities are on the same sublattice and anti-ferromagnetic on the opposite sublattice\cite{saremi2007,brey2007}, decaying as $1/R^3$ as expected \cite{Sherafati.RKKY_graphene_GF}.
We point out that distances here refer to the relative positions of the impurities as illustrated in the inset, and are not to scale (clearly, the distances $x_1$ and $x_2$ are not the same). 

\begin{figure}
\centering
\includegraphics[width=0.48\textwidth]{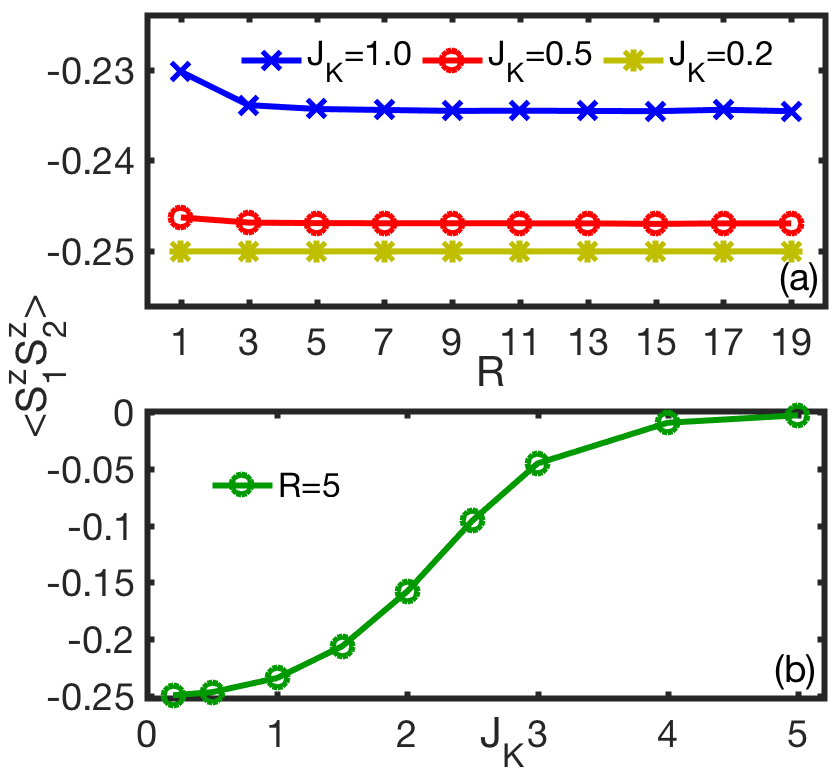}
\caption{(a) Inter-impurity spin-spin correlations for impurities along the $x$-direction at odd distances. Impurities on opposite sublattices are in the free moment regime for small $J_K=0.2$, reflected in a saturation of the correlations (see text). For larger values of $J_K$ correlations become anti-ferromagnetic and do no exhibit any noticeable decay with distance. 
(b) Results at distance $R=5$ as a function of $J_K$. As the interaction increases, eventually correlations get suppressed.
}
\label{figure_afm}
\end{figure}

Figs.~\ref{slg_hf}(c) and (d) show the spin-spin correlations between impurities at half filling. Only the $z$-component of the correlation is shown since the system is $SU(2)$ symmetric. 
Correlations are clearly ferromagnetic for impurities on the same sublattice up to a distance of the order of 15 lattice sites where we see the correlations become very small at some points, corresponding to the onset of Kondo screening. 
For opposite sublattices, instead of a competition between Kondo and RKKY, we see a competition between RKKY and the free moment regime. In Fig.\ref{slg_hf}(c) we show only the results for spins on the same sublattice. When the impurities are on opposite sub-lattices the correlations are identically or very close to the saturation value $\langle S^z_1S^z_2\rangle=-1/4$, and are not shown. In this regime, the magnetic moments are completely decoupled from the conduction electrons and from each other, and the ground state is 4-fold degenerate with spins pointing in either direction. Since we are enforcing spin conservation and $S^z_\mathrm{Tot}=0$, the impurities are always anti-parallel and correlations can only assume the value $-1/4$. We point out that this may always occur in a Kondo box or for a pseudogap density of states for sufficiently small $J_K$, regardless of the position of the impurities. For sufficiently large $J_K$, the impurities will couple anti-ferromagnetically. In Fig. \ref{figure_afm} we show the spin correlations at odd distances. As the interaction increases, eventually the correlations get suppressed, as seen in panel (b) for $R=5$. More interestingly, correlations do not exhibit any decay with distance, indicating that that RKKY is very robust and that Kondo has very little effect in undermining anti-ferromagnetism, at least for small values of $J_K$. These results are telling us that the two impurities are practically decoupled from the bulk forming an almost perfect singlet. Actually this agrees with a perturbative picture in which the conduction electrons introduce and effective interaction between the spins. Unless there is another mechanism competing with the RKKY interaction (such as Kondo), the impurities will form a perfect singlet. The surprising lack of decay may be attributed to the extended nature of the electronic wave-functions near the Fermi level. Certainly this is an interesting problem that deserves further investigation.


\begin{figure}
\centering
\includegraphics[scale=0.33]{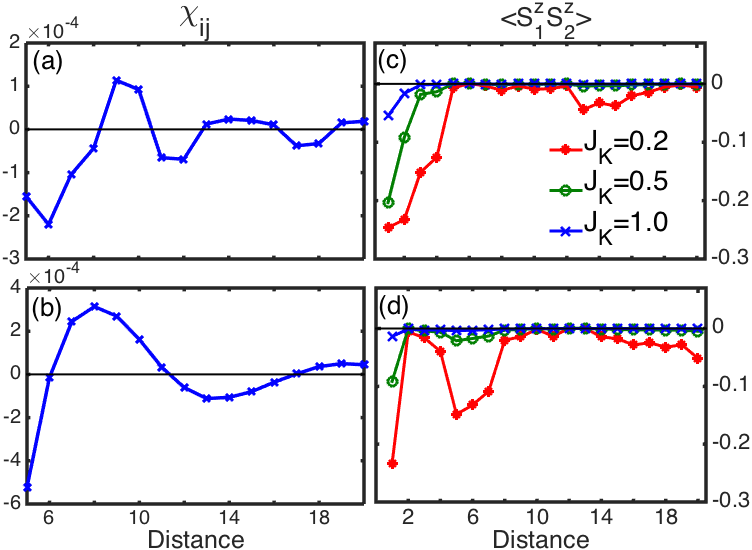}
\caption{Same as Fig.~2 for 13\% hole doped MLG. Vanishing correlations correspond to two uncorrelated Kondo clouds.}
\label{slg_37}
\end{figure}

\begin{figure}
\includegraphics[scale=0.36]{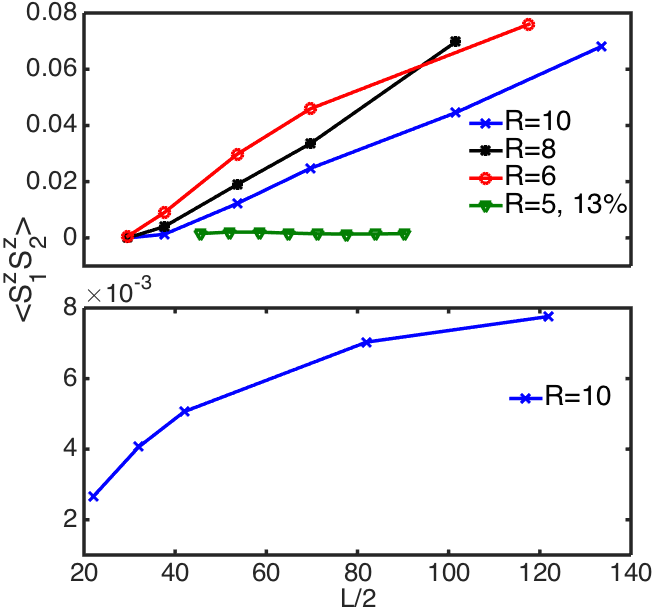}
\caption{Spin-spin correlations for $J_K=0.5$ as a function of system size and different impurity separation for (a) graphene and (b) bi-layer graphene along the x-direction.
The MLG correlations monotonically increase at half filling, indicating a strong trend toward ferromagnetism. Away from half-filling, the Kondo effect is very robust and dominates the physics. In the case of BLG, the correlations smoothly converge and saturate to a finite value.
}
\label{finite_effects}
\end{figure}

On the other hand, when spins are on the same sublattice, they prefer to couple ferromagnetically into an RKKY triplet state, closely resembling the Lindhard function. 
The fact that correlations decay with distance in the ferromagnetic case is only due to the competition with the Kondo interaction and the entanglement with the conduction electrons. 
We have found that these simulations are very susceptible to finite size effects, which are more dramatic at half-filling and for small values of $J_K$. In Fig.~\ref{finite_effects}(a) we show the correlations as a function of chain length at half filling for $J_K=0.5$ for different distances. The correlations grow monotonically without indication of convergence for chain lengths up to $L=280$. 
Despite the apparent lack of convergence, the trend is clear, and indicates that ferromagnetism is very robust.

\begin{figure}
\centering
\includegraphics[width=0.48\textwidth]{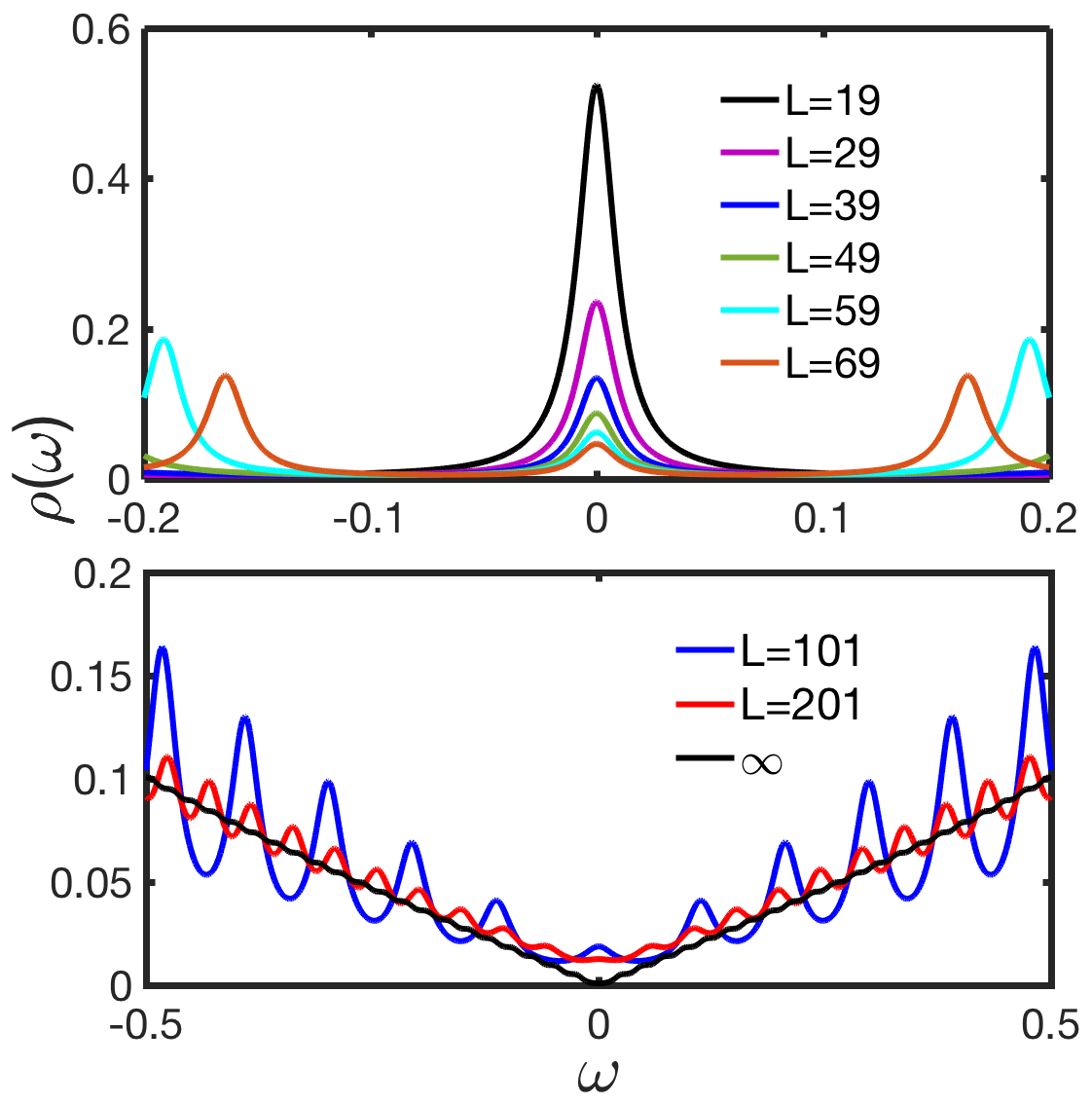}
\caption{(a) Local density of states at the position of one of the impurities for different sizes. The spectrum shows zero energy excitations corresponding to the edge states. (b) Same as (a) but zooming away in energy and for larger system sizes, showing how the pseudogap DOS is recovered.
Notice that the area of the graphene flake is proportional to $L^2$, a relatively large number of carbon atoms.
}
\label{ldos}
\end{figure}

\begin{figure}
\centering
\begin{minipage}{0.48\textwidth}
\includegraphics[width=\textwidth]{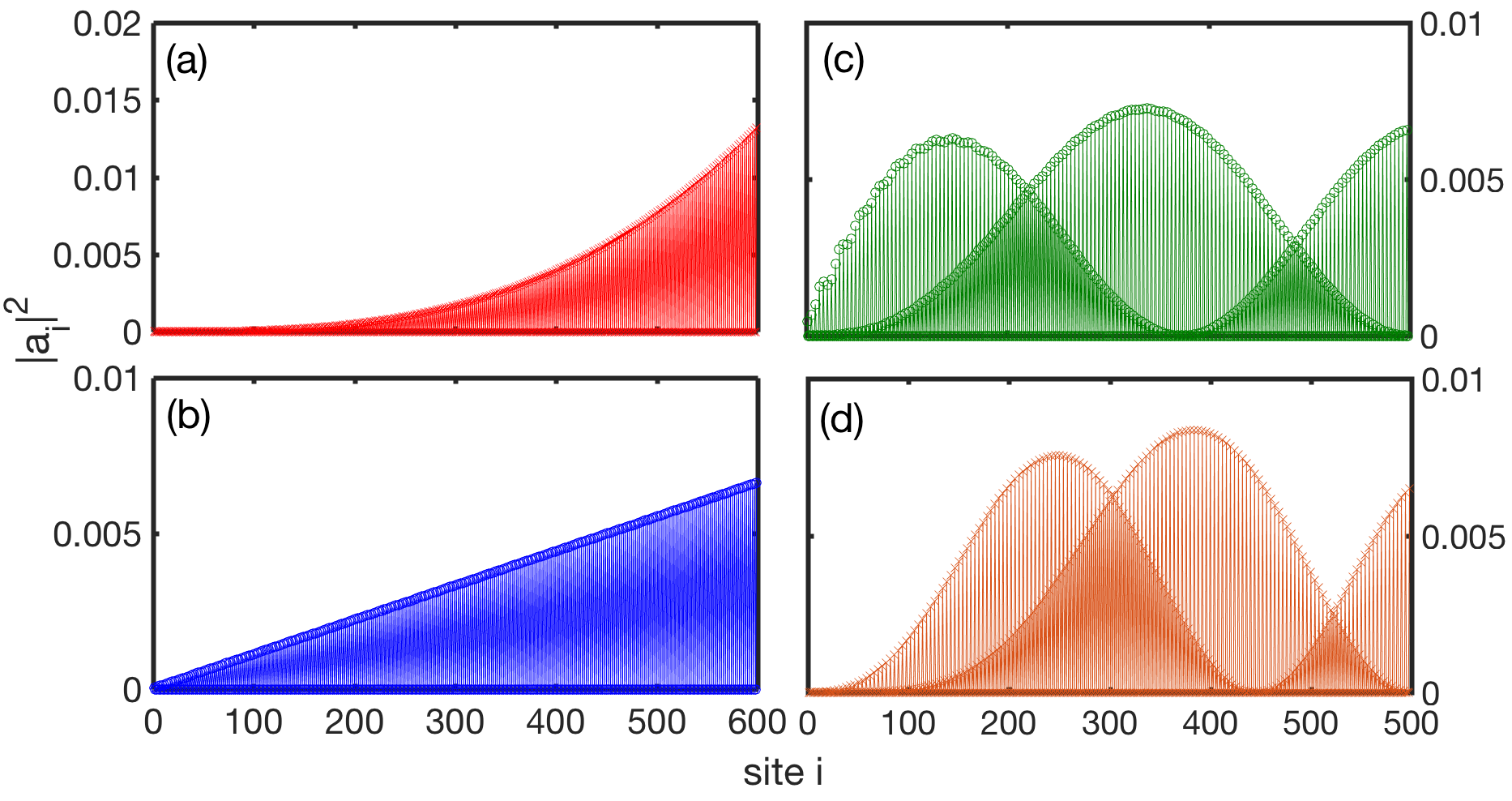}
\end{minipage}
\begin{minipage}{0.48\textwidth}
\includegraphics[width=\textwidth]{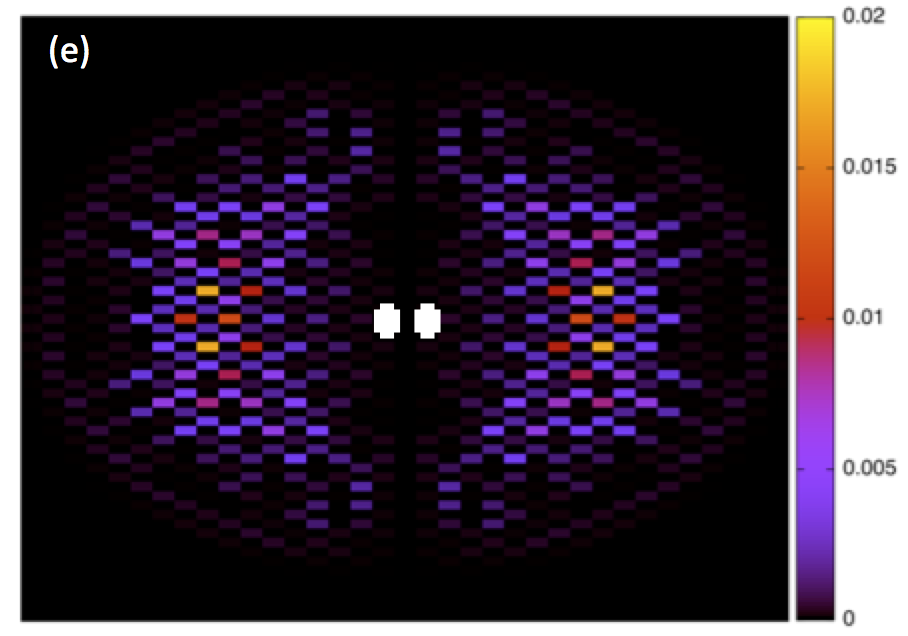}
\end{minipage}
\caption{Wave function amplitude of the zero-energy edge states along the symmetric (a) and antisymmetric (b) chains in the Lanczos basis. The edge modes live only on the same sublattice as the seeds. (c) and (c) show the first excited state along the symmetric and anti-symmetric chains, respectively. Panel (e) shows the square of the wave-function amplitude for the antisymmetric edge state in real space, for $L=30$ and $R=4$. Crosses show the position of the impurities.}
\label{figure_edge}
\end{figure}

We attribute these effects to the presence of zero-energy edge modes. The chains are finite, corresponding to finite flakes with a boundary. These zero-energy modes play a dominant role in small graphene flakes, as seen in the local density of states (LDOS), Fig.~\ref{ldos}.
Due to the symmetry of the orbitals, we find that two impurities on opposite sublattices do not have edge states, and the finite size effects are negligible (not shown). For impurities on the same sublattice, the flake has zig-zag edges. 
In this case we encounter two zero-energy modes:
One of these states decays algebraically down the anti-symmetric chain, while the other one decays linearly along the symmetric channel, as shown in Fig. \ref{figure_edge}(a)-(b). For comparison we also show typical extended wave-functions away from the Dirac points in panels (c) and (d). These results may seem counterintuitive, since in principle one assumes that the zero energy wave-functions decay exponentially from the edge. In fact, this is true for pristine zig-zag edges \cite{Fujita1996}, but it does not apply to a graphene flake of irregular shape as the one considered in this work. Theoretical and experimental studies \cite{Kobayashi2006} show that a combination of different boundary conditions can yield slowly decaying edge states that leak into the bulk, as shown in Fig.\ref{figure_edge}(e) for a typical case. When these wave functions are expressed in terms of the Lanczos basis states, the coefficients decay smoothly and monotonically from the end of the chains. The edges and the bulk electrons compete in order to form a correlated state with the impurities. As the chain length is increased, the weights of the zero energy states near the impurities and their spectral weight in the LDOS decrease, and the bulk electrons win, enhancing the RKKY interaction and causing an increase in the correlations. At a sufficiently long chain length, the effects should become negligible.

\begin{figure}
\centering
\includegraphics[width=0.49\textwidth]{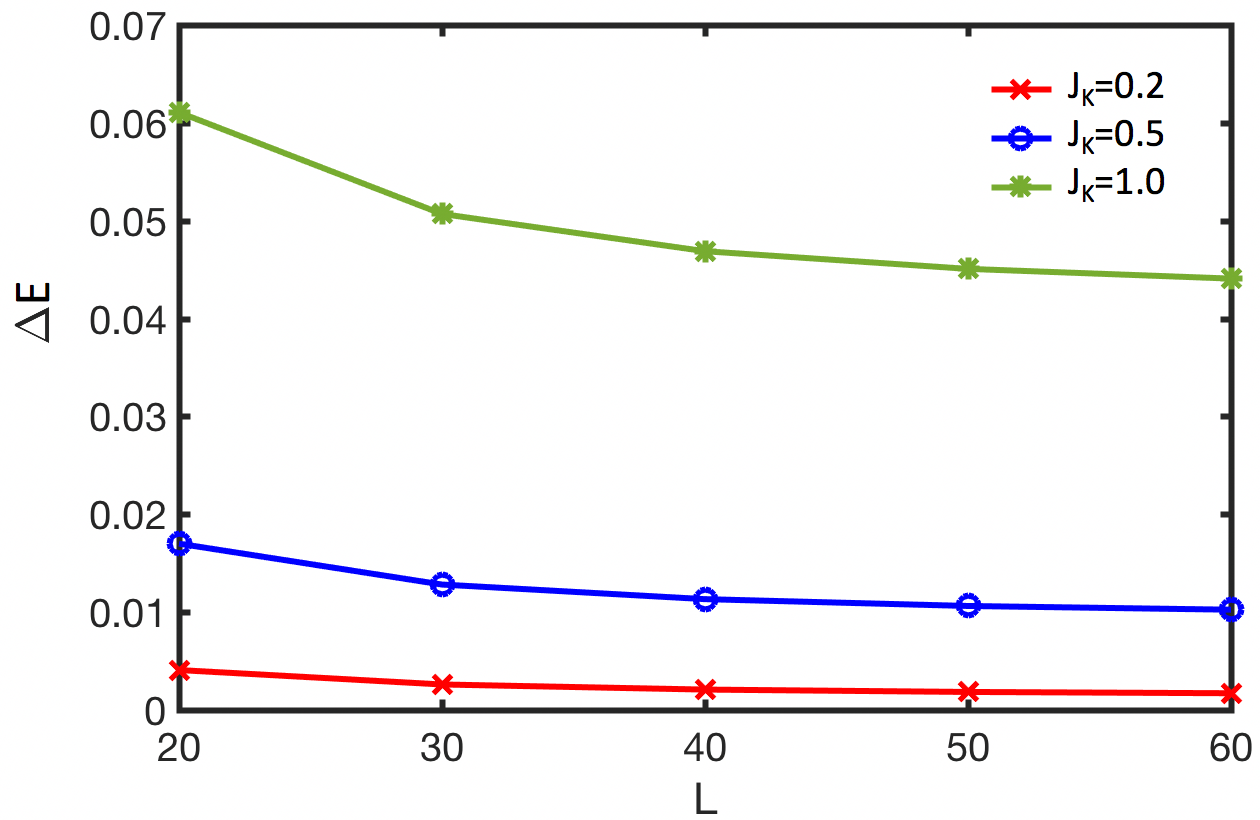}
\caption{Energy gain $\Delta$ (as defined in the text) for a single impurity as a function of system size $L$. These values should reflect an estimate for the Kondo temperature $T_K$. For small graphene fragments we expect edge states to play a more dominant role. $\Delta$ decreases and plateaus when the impurity decouples from the edges and the physics is dominated by the bulk.}
\label{energy_gain}
\end{figure}

The influence of the edge states is also observable in the single impurity case. In Fig.\ref{energy_gain} we plot the energy gain after coupling the impurity to graphene, defined as \cite{Varma1976}:
\[
\Delta E = |E_0(J_K)-E_0(J_K=0)|.
\]
This quantity is an estimate of the Kondo temperature $T_K$, or the energy needed to break the Kondo singlet. As expected, it increases with $J_K$, but it {\it decreases} with $L$. The reason is that for small systems, when the impurity interacts with the edge states, there is a relatively large spectral weight at the Fermi energy. When $L$ is large, the inter-level spacing no longer plays a role and $\Delta E$ plateaus at a value that is independent of the system size, indicating a property of the bulk.
This is a dramatic reflection of the significance of edge modes in finite systems.

If the Fermi level is moved away from the Dirac point by doping the system, the physics changes in a notable way. From the RKKY expression, Eq.(\ref{lindhard}), the Fermi wave vector determines the wavelength of oscillations in the RKKY interaction as shown in Figs.~\ref{slg_37}(a) and (b) for the 13\% (measured from the Dirac point) hole doped case. The difference in wavelength of oscillations is due to the fact the units of distance differ in the two directions. While calculations were also done at other fillings and for electron doping, results for just one value are presented. Electron doped results are identical due to the particle hole symmetry. Figs.~\ref{slg_37}(c) and (d) show the numerical results for the spin correlations for different values of $J_K$. As the coupling is increased, the correlations undergo a crossover from the free moment regime, to anti-ferromagnetic RKKY, and then to Kondo, with ferromagnetism being completely absent. 

\begin{figure}
\centering
\includegraphics[scale=0.3]{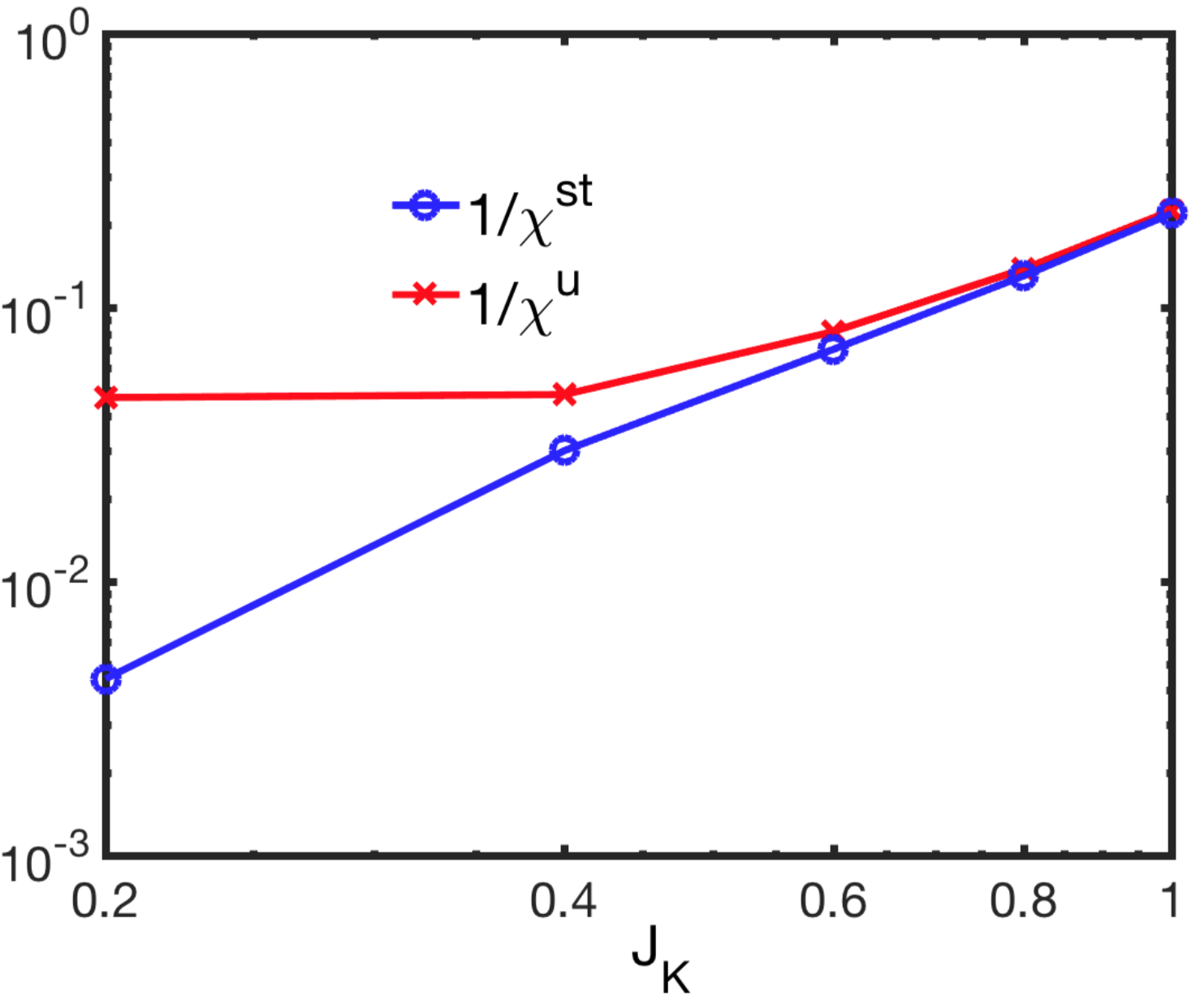}
\caption{Staggered and uniform susceptibilities as a function of $J_K$ for distance $R=4$ and 13\% hole doping. Both quantities become equal when the Kondo regime is reached. Results are obtained by applying a small magnetic field $h=10^{-4}$ on the localized spins.
}
\label{chi}
\end{figure}

When the impurities are in a Kondo state, the correlations are identically zero. This situation was extensively discussed in Ref.~\onlinecite{Allerdt.Kondo}. Each impurity is fully screened by the conduction electrons and they form two independent Kondo singlets. We determine this by analyzing the staggered and uniform magnetic susceptibilities in Fig.~\ref{chi}, and verifying that they are identical, and equal to the single impurity result. 
The effect is even more dramatic as the Fermi level is moved further away from the Dirac point (not shown), with correlations vanishing after just a few lattice spacings. Results at other fillings and system sizes also do not show any indication of ferromagnetism. 
Finite size effects are also notoriously weaker (practically ignorable) than the half-filled case, as seen in Fig.~\ref{finite_effects}(a), since we are far from the particle-hole symmetric point where level spacing and edge modes become irrelevant.

\begin{figure}
\centering
\includegraphics[scale=0.41]{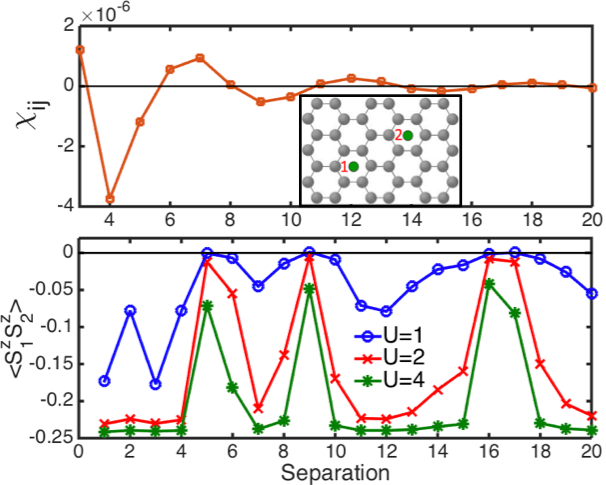}
\caption{Lindhard function (top) and spin-spin correlations (bottom) for two impurities at the center of hexagons in $5\%$ hole doped graphene and parameters: $V=0.2$, $\epsilon_{d}=E_F-\frac{U}{2}$. Distances are measured along the zig-zag direction as shown in the inset.}
\label{plaquette_correlations}
\end{figure}

 We next consider the case of ad-atoms sitting at the center of the hexagon. To study this problem we use the Anderson model, Eq. (\ref{hamiltonian2}). 
Here we consider impurities spaced along the zig-zag/diagonal direction. Using the same method as above, we are again able to measure the correlations between the impurities. We assume $\epsilon_d=E_F-U/2$, and we take $t'=0.2$. Fig.~\ref{plaquette_correlations} shows spin correlations  as well as the perturbative result for graphene at $5\%$ hole doping. The results display similar behavior as the previous cases away from half filling, with dominant anti-ferromagnetic correlations, and a crossover from the free moment to the Kondo regime for decreasing $U$ (increasing $J_K \sim t'^2/U$). At half-filling, the impurities are always in the free-moment regime, as also found for the single impurity case in Ref.~\onlinecite{Yunoki.Block}.
We emphasize that perturbative RKKY considerations based on the Lindhard function miss all of this important interplay with Kondo physics and clearly depart from the numerical results.

\subsection{Bi-layer graphene}
Bilayer graphene is composed of two layers of graphene stacked on top of each other. We study two forms of BLG: symmetrically stacked such that the sublattices of each layer coincide, and `Bernal' stacked, with sublattice A directly above sublattice B. 
The two forms of BLG have different band structures \cite{Partoens.From_graphene,McCann.Landaulevel} that can each be approximated by a four band model. The Bernal structure has a parabolic dispersion near the Fermi level that can be further reduced by considering just two bands, since the other two are separated by an energy on the order of the interlayer hopping $t'$. In previous work \cite{Hwang2008}, it was shown that this two-band problem yields a trivial RKKY interaction between impurities. However, if one considers linear contributions, second neighbor interactions\cite{Partoens.From_graphene}, or values of $J_K \gtrsim t$, these assumptions are no longer valid, and a more general four-band model is required, as the one used here.

\begin{figure}
\includegraphics[scale=0.31]{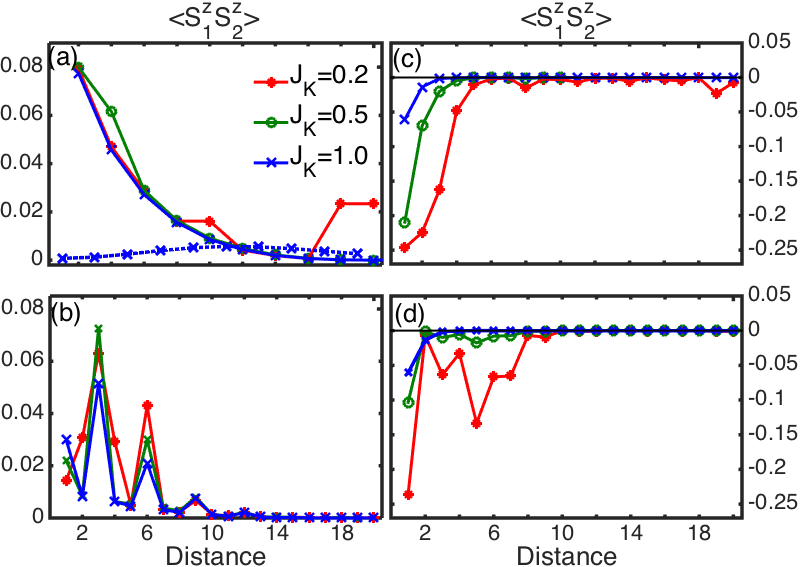}
\caption{Spin-spin correlations for BLG at half-filling along the (a) $x$-direction and (b) $y$-direction on the same layer. Dotted line in panel (a) corresponds to spins on opposite layers. Missing data points indicate that spins are in the free moment regime. Panels (c) and (d) show results for $13 \%$ doping. The dashed line in (b) represents impurities on opposite layers for $J_K=1$.
}
\label{blg}
\end{figure}

We focus on the symmetric stacking, and results for the Bernal stacking look qualitatively very similar. The hopping within a layer is taken to be the same as pure graphene, while for the one between layers we use $t'= 0.1t$. Note that unlike graphene, BLG has a small but finite density of states at the Fermi level. The Lindhard function along the $x$ and $y$ directions (not shown here) is qualitatively similar to the one for MLG, with the sign of the interactions reversed when impurities are on opposite layers. 
Numerical non-perturbative results at half-filling are shown in Fig.~\ref{blg}(a) and (b). The correlations on a single layer look identical to those for graphene, but decay faster due to the increased DOS near the Fermi point, and also due to the increased dimensionality which interpolates between 2D and 3D. However for the BLG case, impurities on opposite layers along the $y$-direction are in the free moment regime, or anti-ferromagnetic (results look similar to those in Fig.\ref{figure_afm} and are not shown), while along the $x$-direction they are weakly coupled ferromagnetically if on the same sublattice. Away from half-filling, we find, same as for graphene, that the correlations completely depart from the expected RKKY behavior and ferromagnetism is absent, as shown in Fig.\ref{blg}(c) and (d) for 13\% doping. We find the same qualitative behavior at other doping densities. 

\section{Conclusions}
The cooperative many-body behavior of quantum impurities in 2D materials, such as graphene and bilayer graphene, is complex and defies intuition, with a non-trivial competition between screening and magnetism. In addition, due to the small density of states at the Fermi level, impurities may not couple to the conduction electrons at all, behaving as free moments. 
Our numerical non-perturbative results show that indirect exchange at half-filling is quite well described by the perturbative RKKY interaction in the ferromagnetic case when both impurities are on the same sublattice. However, anti-ferromagnetism in the undoped half-filled case only becomes a dominant feature for $J_K$ larger than a critical value and impurities remain in the local moment regime otherwise. 
Recent experiments with hydrogen atoms on MLG \cite{Gonzalez-Herrero2016} show vanishing coupling between substrate and spins when they are placed on opposite sublattices, in agreement with our results, that predict free moments. DFT calculations\cite{Gonzalez-Herrero2016} indicate that this effect may be due to single particle physics, and explained by the way electrons occupy different orbitals once one accounts for the chemistry of the problem. 
This experimental example  can be described in terms of $s$ and $p$ orbitals and is weakly interacting. The physics departs from our strongly interacting regime and is probably simpler and well described by DFT and/or perturbative techniques.
Our work applies mainly to the case of extrinsic magnetic impurities in graphene, the corresponding situation for defect or vacancy induced intrinsic magnetism in graphene is more complex.
We point out, however, that the lack of spin signature would also occur in the case of free moments, in which case the impurity would be transparent, as recently observed in Ref.~\onlinecite{Stenbrecher2016} for Ho atoms adsorbed on Pt(111). 

At finite doping, the departure from the RKKY theory is more dramatic: ferromagnetism is completely absent, in contrast to the result in Ref.~\onlinecite{brey2007}, and the physics is completely dominated by a competition between anti-ferromagnetism and Kondo. 

For bilayer graphene, impurities on opposite layers remain in the free moment regime, unless the interaction $J_K$ is increased to values of the order of half the hopping $t$. Away from half filling, in the doped situation, oscillations between ferro and anti-ferromagnetic phases are absent, significantly departing from the expected $1/R^2$ decay found in Ref.~\onlinecite{Hwang2008}. These results highlight the importance of the correlations in this problem and the failure of perturbative approaches in studying these phenomena. In addition, they illustrate the relevance of the band structure, the interference effects of the electronic wave function on the lattice \cite{Feiguin.Lanczos,Mitchell2015,Schwabe2015}, and the presence of edge states at half-filling in small flakes. 

This work indicates a route toward realizing a dilute anti-ferromagnet in graphene, and emphasizes the key importance of the non-perturbative interplay between Kondo and RKKY physics in determining adatom-induced graphene magnetic properties. We hope that our exact calculations of adatom-induced graphene magnetic properties will motivate experimental studies of MLG and BLG magnetism.
 
\begin{acknowledgments}
We are grateful to C.~F. Hirjibehedin and S. Kar for useful discussions. AA and AEF acknowledge U.S. Department of Energy, Office of Basic Energy Sciences, for support under grant DE-SC0014407. SDS is supported by LPS-MPO-CMTC. 
\end{acknowledgments}

\bibliographystyle{apsrev}

\end{document}